\documentstyle[12pt]{article}
\newcommand{\beq}{\begin{equation}}
\newcommand{\eeq}{\end{equation}}
\newcommand{\ds}{\displaystyle}
\newcommand{\lng}{\ln g}
\begin{document}  \begin{flushright} PITHA 97/40 \\ gr-qc/9710032
\end{flushright}
   \begin{center} {\LARGE  Canonical  Quantum
 Statistics
  \\[0.2cm] of  Schwarzschild Black Holes \\[0.2cm] and
  Ising Droplet Nucleation\\[0.3cm] }
{\Large H.A. Kastrup\footnote{E-Mail: kastrup@physik.rwth-aachen.de} \\
Institute for Theoretical Physics, RWTH Aachen \\[0.2cm] 52056 Aachen,
 Germany}
\end{center} \vspace*{0.5cm}
  {\large \bf Abstract} \\[0.1cm] \\ Recently it was shown[1] that the
   imaginary
  part of the canonical
  partition function of  Schwarzschild black holes with an energy spectrum
  $ E_n = \sigma \sqrt{n}\, E_P\;, n=1,2,\ldots,$ has properties
   which - naively
  interpreted - leads to the expected unusual thermodynamical properties of
  such black holes (Hawking temperature, Bekenstein-Hawking entropy
   etc.). \\
  The present paper interprets the same imaginary part in the framework
  of droplet nucleation theory in which the rate of transition from a
  metastable state to a stable one is proportional to the imaginary part
  of the canonical partition function. The conclusions concerning the
  emerging thermodynamics of black holes are essentially the same as before.
  \\ The partition function for  black holes with the above spectrum
    was calculated exactly recently[1]. It is the same as that of the
     primitive
  Ising droplet model for nucleation in 1st-order phase transitions in 2
  dimensions. Thus one might learn about the quantum statistics of black
  holes by studying that Ising model, the exact complex free energy of which
   is presented here for negative magnetic fields, too.

  \newpage
   \section{Introduction}In a recent paper\cite{ka1} I discussed properties
   of the canonical partition function  for a quantum energy spectrum
   \beq E_n = \sigma \sqrt{n}E_P\;, n=1,2,\ldots,~~ E_P=
    c^2 \sqrt{c\, \hbar/G}~,~
   \sigma=O(1)~~, \eeq where the n-th
   level has the degeneracy \beq d_n = g^n~,~g>1~~. \eeq The main
    interest in this
   spectrum comes from the many, differently justified, proposals
    (see the corresponding quotations
   in ref.\ [1]) that  a quantized Schwarzschild black hole might have such
   a spectrum!
   The associated canonical  partition function
    \beq Z(t,x) =
   \sum_{n=0}^{\infty}e^{\ds n t}e^{\ds -\sqrt{n}x}~,
   ~~t= \ln g~,~
    x= \beta \sigma E_P~~, \eeq converges only for $|g| \le 1$,
     whereas one is interested
    in its properties for $g>1$. However, the function $Z(g,x)$
    can be continued analytically into the  complex $g$-plane by means
    of the integral representation[1]
    \begin{eqnarray} Z(g=e^t,x)&=&
   \frac{x}{2\sqrt{\pi}}\int_1^{\infty}du
  \frac{\ds e^{\ds -x^2/(4\ln u)}}{\ln^{3/2}u}\frac{1}{u-g}~~ \\ &=&
 \frac{x}{2\sqrt{\pi}}\int_0^{\infty}\frac{d\tau}{\tau^{3/2}}
  e^{\ds -x^2/(4\tau)}\frac{1}{1-e^{\ds (t-\tau)}}~,
   \end{eqnarray}  which exhibits a branch cut of $Z(g,x)$ from $g=1$ to
   $g= + \infty$. Approaching the cut from above yields a complex $Z$, with
   an imaginary part
   \beq Z_i(t,x) =
  \frac{\sqrt{\pi}x}{2t^{\ds 3/2}} e^{\ds -x^2/(4t)}~~ \eeq
   and the principal value integral \beq
  Z_r(t,x)= \mbox{p.v.} \frac{1}{\pi}\int_0^{\infty}d\tau\, Z_i(\tau,x)
  \frac{1}{1-e^{\ds t-\tau}}~  \eeq
 as its real part. \\  Amazingly one gets essentially all the expected
thermodynamical properties of a black hole, including the Hawking
 temperature,
if one - naively - uses the imaginary part $Z_i$ above for their derivation.
This is strange and provocative and needs further analysis. If such an
 approach
can be substantiated, one will have an indication which quantum states
of a black hole could be responsible for its unusual thermodynamical
properties! \\In the following I shall try to put the properties mentioned
into a certain perspective and especially discuss their relationship
 to other
approaches (which was not done in ref.[1]), namely the euclidean
 path integral (semiclassical) quantisation
of black holes by Hawking and others and the theory of droplet nucleation
associated with metastable states. \\ The paper is organized as follows:
In chapter 2 the properties of the model are briefly summerized and its
theoretical and physical potentials and limitations sketched. Parts of it
recall well-known material which serves to have the discussion
more or less self-contained. Chapter 3
discusses the relationship of the model to Hawking's euclidean path integral
approach to the thermodynamics of  black holes, where he also gets a purely
imaginary partition function\cite{ha1}!
 Chapter 4 deals with the nucleation of droplets
occuring when a system changes from a metastable state to a stable one, e.g.
in 1st-order phase transitions or in the decay of "false" ground states.
Langer\cite{la1} was the first one to relate the imaginary part
 of the canonical partition
function to the transition rate of such processes. Later Gross,
Perry and Yaffe\cite{gr} interpreted the imaginary part of the semiclassical
euclidean path integral - i.e.\ the canonical partition
function - for the Schwarzschild black hole ("instanton") in this
 context, thus
giving a different interpretation of the same quantity Hawking employed. \\
In chapter 5 the relation of the partition function (3) to that of the
 primitive Ising droplet
 model for nucleation in 2 dimensions is discussed and it is very
  easy to see
that both are exactly the same.
 The integral representation (4)
therefore provides an exact free energy expression for this model,
 too, which as
far as I know, is new\cite{mu}. Chapter 6 presents some conclusions.
\section{Scope of the model}In the background of trying to push the quantum
analysis - including its quantum statistics - of the simple isolated
Schwarzschild gravitational system as far as possible are the following
considerations: The quantized radiation emitted by a "classical"
Schwarzschild black hole has a thermal (Planck) distribution governed by
the (Hawking) temperature\cite{ha2}
\beq
   k_B T_H \equiv \beta^{-1}_H = \frac{c^3\hbar}{8 \pi G M}~, \eeq
    where $M$ is the mass
   of the black hole. Assuming that this temperature of the radiation
   arises from its thermal contact with the black hole of the same
   temperature, the immediate question was - and still is -, how does the
   black hole gets this temperature? As the  temperature (8) is
   proportional to Planck's constant the quantum theory of gravity
   has to play a decisive role in its understanding. Especially one would
   like to identify the microscopic quantum gravitational degrees of freedom
   which yield the corresponding quantum statistics and its macroscopic
   thermodynamics. \\ Many attempts have been made to achieve this
    and I shall
    mention only a  few
   of them in the following. The most active and enthusiastic reseach in
   this area presently takes place in connection with string theory. I
   shall say nothing about that in the following and refer to the very
   recent review by Horowitz\cite{ho}. \\ In view of the lack
   of a generally accepted quantum theory of Einstein's General
    Relativity it
   seems difficult to extract those microscopic gravitational
   degrees of freedom within this framework. However, the situation is not
   quite es hopeless as it might appear. The reason is that
    Einstein's theory
   for a rotationally symmetric isolated gravitational system,
   Schwarzschild gravity, can consistently be quantized, namely by first
   identifying its classical observables in the sense of Dirac by solving the
   constraints associated with the gauge (diffeomorphims) degrees of freedom
    and then quantizing the remaining
  gauge-independent physical degrees of freedom  afterwards. This has been
  done by Thiemann and myself\cite{ka2} in the framework of Ashtekar's
   formalism
  and briefly afterwards by Kucha\v{r}\cite{ku} in the geometrodynamical
  framework, the results being equivalent. They have been briefly summerized
  in ref.\cite{ka3}. \\ The essential point is that this system classically
  has just
  one canonical pair of (Dirac) observables, namely its mass $M$ and as the
  canonically conjugate quantity of $M$ a time functional $ T[g;\Sigma]$
  (of the metric $g$) which
  describes the  difference $\tau_+ - \tau_-$ in proper time $\tau$ of two
  observers at the two (asymptotic) ends of the 1-dimensional spacelike
   hypersurface
  $\Sigma$ which, together with the 2-dimensional spheres $S^2$, provides
  a time slicing of the 4-dimensional manifold. Formally the quantity
  \beq \delta = T[g;\Sigma]-(\tau_+ -\tau_-) \eeq is  the constant Dirac
  observable. \\ Thus, for an observer at the asymptotic end
 where $r \rightarrow + \infty$ on the Schwarzschild
  manifold
 with the time
  $\tau \equiv \tau_+$ there are - in the framework of this
   model - only two quantities
  available in order to describe the system: its Mass $M$ and his proper time
  $\tau$! Everything else has to be expressed by them (and some fundamental
  constants like $G,c,\hbar,k_B$). Examples are: the Schwarzschild radius
  $R_S = 2MG/c^2$ or any multiple thereof, the area $A=4\pi R^2_S$
   of the horizon, but also any
  time interval $\Delta$ associated with the system, i.e.\ $\Delta = \gamma
  R_S/c$, where $\gamma$ is some number, e.g. $\gamma=4\pi$ for the time
  period $\Delta_H$ of the euclidean section\cite{ha1} of the complex
   Schwarzschild
  manifold. \\ Having eliminated the (infinite) gauge degrees of freedom
  classically one can now quantise the remaining physical degrees
  of freedom[8-10]. The extremely simple Schr\"odinger equation for
   this system,
  \beq i\hbar \partial_{\tau} \phi(\tau) =Mc^2 \phi(\tau)~, \eeq
 has the plane wave solutions \beq \phi(M,\tau)= \chi(M)e^{\ds
 -\frac{i}{\hbar}Mc^2\tau}~, \eeq where $M$ is continuous and $>0$.
 So at first sight there is no such spectrum like (1). However, one
 can get it by changing the boundary conditions. \\ My doing so in ref.[10]
 was stimulated by the paper of Bekenstein and Mukhanov\cite{be1}, in
 which they discussed properties of the spectrum (1) with
 $\sigma^2=\lng/(4\pi), g=2$. Already in 1974 Bekenstein\cite{be2} discussed
 this spectrum as arising from a Bohr-Sommerfeld type quantisation of the
 area of the horizon ($A \propto n\hbar$). Since then many authors have proposed
 such a spectrum (see refs.\ [3-23] in ref.[1]). \\
 One knows from elementary quantum mechanics how to make the continuous
 momentum in plane waves discontinuous by periodic boundary conditions
 in space. Applying corresponding boundary conditions in time (with period
 $\Delta$) to the
 plane waves (11) yields \beq c^2M\Delta = 2\pi \hbar~ n~,~ n=1,2,
 \ldots~.
   \eeq
 Introducing such boundary conditions is easier than justifying them by
 physical arguments. In any case it means that the system is represented
 by the plane waves only for the time $\Delta$ after which it is abruptly
 terminated, e.g., by forming of the horizon. This may happen, e.g.,
  because the
 state described by the plane wave is a metastable one - see below.
 There are several physcal time scales
 $\Delta=\gamma R_S/c$ associated with a black hole  all of which[10]
  have $\gamma
 \sim O(1)$, so that \beq E_n = \sqrt{\frac{\pi}{\gamma}}
 \, \sqrt{n}\, E_P \equiv \sigma
 \sqrt{n}\, E_P~~. \eeq
\section{Euclidean path integral as
 canonical partition function of a Schwarzschild black hole}First I want
  to recall how
Hawking's path integral approach to the canonical partition function in a
semiclassical approximation leads to a purely imaginary partition function,
too: \\ Shortly after Hawking's
discovery that the quanta emitted by a collapsing black hole have a thermal
distribution governed by the temperature (8)
 Gibbons and Hawking\cite{gi1} employed the
relationship between the canonical partition function of a system
 and the corresponding
euclidean path integral  in order to tackle the canonical quantum
statistics of a black hole itself. For pure gravity that partition
 function $Z$
is formally given
by \beq  Z= \int D[g] e^{\ds -I_E[g]/\hbar}~,~~ I_E=
\frac{1}{16\pi G}\int d^4x
R(g)(g)^{1/2}+ \mbox{surface terms},\eeq where $I_E$ is the
 euclidean action of pure gravity
 (for more details  of the path integral approach to quantum gravity
 see the reprint collection in ref.[2]). In principle one has to take into
 account all field configurations ${g_{\mu \nu}}$ which
  are periodic in euclidean time with period $\hbar \beta$ and which obey
  appropriate spatial boundary conditions. In practice the path
   integral has been
  evaluated only semiclassically: One puts \beq g_{\mu \nu} =
  \stackrel{0}{g}_{\mu \nu} +
  h_{\mu \nu}~, \eeq where $\stackrel{0}{g}_{\mu \nu}$ is a
   classical solution
  of the euclidean field equations compatible with the boundary conditions
  just mentioned
   and the $h_{\mu \nu}$ denote (small) quantum
  fluctuations around the classical solution. Inserting (15) into the
   action (14)
   and keeping only terms up to order $O(h^2_{\mu \nu})$ gives
  the semiclassical approximation  \beq I_E^{scl} =
  I_E^{cl}[\stackrel{0}{g}]+I_E^{(2)}~,~~ I_E^{(2)}=\int
  d^4x(\stackrel{0}{g})^{1/2}(hA(\stackrel{0}{g})h)~,  \eeq
  where $A(\stackrel{0}{g})$ is a second order differential operator
  depending on the classical metric $\stackrel{0}{g}_{\mu \nu}$. \\
Gibbons and Hawking evaluated $I_E^{cl}$ for the euclidean Schwarzschild
metric and obtained \beq I_E^{cl}[\mbox{\small Schwarzschild}] = \frac{1}{2}
\beta_H Mc^2~,
 \eeq where the period $\hbar \beta_H$ follows from the requirement that the
 euclidean Schwarzschild section has no conical singularities.
  The associated partition function \beq Z^{cl}_S =
e^{\ds -\beta_H Mc^2/2}= e^{\ds -E_P^2 \beta_H^2/(16\pi)} \eeq gives
 the desired thermodynamics of black holes,
especially the Bekenstein-Hawking entropy \beq S_{BH}/k_B=
\frac{A}{4l_P^2}~,~ l_P^2=G\hbar/c^3~. \eeq The approach appears to become
"unconventional", if one includes the term $I_E^{(2)}$ for which the path
integral (14) is a Gaussian one which - formally - is proportional to the
inverse square root of the product of the eigenvalues of the operator $A$
(=$(\mbox{det}(A))^{-1/2}$). For
the Schwarzschild case the term $I_E^{(2)}$ has been evaluated by Gibbons and
Perry\cite{gi2}. A crucial point is that the operator $A$ has just one
negative eigenvalue[4] which contributes a factor $i$ to the partition
function. The combined contribution to the partition function $Z^{scl}_S$
of the
 classical euclidean Schwarzschild
solution and the fluctuations around it
is given by \cite{gr,gi2} \begin{eqnarray} Z^{scl}_S
 &=& iZ_g Z_{GHP}~, \nonumber \\ Z_{GHP}&=& \frac{1}{2}
\frac{V}{64 \pi^3 l_P^3}
(a\beta_H)^{\ds 212/45}e^{\ds - E_P^2\beta_H^2/(16\pi)}~,~
 Z_g =e^{\ds \pi^2 V/(c^3\hbar^3\beta_H^3)}~,  \end{eqnarray} where
 $V$ is the
spatial volume of the system and $a$ is a "cutoff" associated with the
$\zeta$-function regularization of $\mbox{det}(A)$. (The index "GHP"
 stands for
"Gibbons, Hawking, Perry".) The factor $Z_g$ is the same as one would obtain
for flat space $(\stackrel{0}{g}_{\mu \nu}= \delta_{\mu \nu}$). It
 represents the gas of free thermal gravitons surrounding the black hole.
 \\ Except for the spatial
volume factor the structure of $Z_{GHP}$ in eq.\ (20) is very similar to
 that of $Z_i$ in eq.\ (6) above: If $\sigma^2=t/(4\pi)$ - see ref.\ [1] -
 then the  exponentials in both cases are equal for $\beta=\beta_H$. The
 factors with the powers of $\beta$ - which in eq.\ (20) comes from
short-distance fluctuations around the classical solution - are different in both
  cases. However this is not
 surprising because both approaches are so different. These factors may give
 rise to logarithmic corrections[1] to the Bekenstein-Hawking entropy
  (19) which
 is dominated by the exponential. I shall come back to this in the next
 chapter. \\
 Hawking has argued[2, ch.\ 15.8] that the partition function (20) has to be
 imaginary in order for the density of states $N(E)$ in the Laplace
  transforms \beq
 Z(\beta)=\int_0^{\infty}dE N(E) e^{\ds -\beta E}~,~~
  N(E)=\frac{1}{2\pi i}\int_{-i\infty}^{+i\infty}d\beta
   Z(\beta)e^{\beta E}\eeq
 to be real if $Z^{cl}$ from eq.\ (18) is inserted into the last integral and
 which exists only if the contour of integration is rotated by $\pi/2$!
Because of this Hawking has concluded\cite{ha1} that the canonical
 ensemble is
inappropriate for the description of black holes. \\ Recently Hawking and
Horowitz have discussed\cite{ha3} a more refined version of the
 relationship between
the classical euclidean gravitational action (14) - especially
 its boundary terms- and the entropy of the system. This work is
  related to
 that of York and collaborators\cite{yo} who suggested to improve
  the approach
 of Gibbons and Hawking[13] and Gross, Perry and Yaffe[4] by employing a more
 microcanonical framework. Related is also the work of Iyer and
  Wald\cite{Iy}.
\\ In any case, the imaginary
parts $Z_{GHP}$ of eq.\ (20) or $Z_i$ of eq.\ (6), respectively,
 do seem to yield
interesting though unconventional thermodynamical properties, if one treates
them as if they were "normal" partition functions. As they imply
 negative heat capacities they signal instabilities of the system. \\
This leads us to a closely related interpretation of the imaginary part of a
partition function, namely of being associated with the decay of a metastable
state into stable one in the form of nucleation: \section{Imaginary part of
partition functions  and droplet nucleation} In a series of
 very influential
papers\cite{la1} Langer discussed the imaginary part of  partition functions
for systems with a 1st-order phase transition where the transition from
a metastable state to a stable one is initiated by the "nucleation" of
droplets. Langer suggested that the nucleation transition rate is
 proportional
to the imaginary part of the partition function. Coleman and Callen\cite{co1}
discussed this approach in the context of instanton solutions in
 the euclidean
version of otherwise lorentzean field theories and their role for the
decay of "false" (metastable) vacua. The methods involved have found wide
applications in different fields of physics\cite{rev}. \\ The main
 procedure is
very similar to the one described above for the euclidean section of the
Schwarzschild solution: The instantons (or "bounces") are classical
 solutions of the
"euclideanized" field equations satisfying appropriate boundary conditions
of a certain path integral. The quantum (or thermal) fluctuations around
these classical solutions leave the system stable if all the eigenvalues
of the second-variation operator $A$ (see above) are positive. However, if
one of the eigenvalues is negative, then the classical solution is
 not related
to a
minimum of the euclidean action integral, but only to a saddle point
 and the fluctuations can drive the system
from its metastable state into a  more stable lower one. \\ The transition
rate is essentially determined by  $\Im(Z^{scl})/Z_0$, where $Z_0$ is the
real part of
 the canonical
partition function (which represents
 the metastable state)
 and by the absolute value of the negative
eigenvalue. \\
It is this framework in which Gross, Perry and Yaffe\cite{gr} interpreted
the contribution (20) of the euclidean Schwarzschild solution ("Schwarzschild
instanton"): \\
 Eq.\ (20) represents the "1-instanton" contribution to the partition
  function.
 The next step is a dilute-gas approximation: Neglecting the "interactions"
 between the instantons, the contribution of N of them to the grand partition
 function $Z_G$ is  \beq Z_g\frac{1}{N!}(iZ_{GHP})^N~  . \eeq Summing
  over all $N$
 gives \beq Z_G = Z_g e^{\ds iZ_{GHP}} \eeq and the grand canonical
  potential
 \beq \Psi(\beta, V, \alpha) = \ln Z_G = \beta p V~,~~ d\Psi =
  -U d\beta + p \beta
 dV - \bar{N}d\alpha~, \eeq
 ($\alpha \equiv - \mu \beta, \mu:$ chemical potential,
   $p$: pressure) here has the form \beq \Psi =
  \ln Z_g + i Z_{GHP}~, \eeq whereas the general structure is[3,19]
  \beq \Psi = \ln Z_0 + i\Im(Z^{scl})/Z_0~. \eeq Because of the special form
  of $Z^{scl}$ in (20) the factor $Z_g$, representing the metastable graviton
  gas (i.e.\ $\ Z_g = Z_0$) drops out in eq.\ (26)! \\
  According to Langer\cite{la1} (and
 others\cite{co1,rev}) the rate $\Gamma$ of transition per unit
  volume from the metastable state (here the
 graviton gas) to the stable state (here the black hole) is given by \beq
 \Gamma = \frac{|\kappa|}{\pi}\Im (\psi)= \frac{|\kappa|}{\pi}
 \Im(Z^{scl}/Z_0)~,~~ \psi =\Psi/V~~,\eeq
  were $\kappa$ is the single
 negative eigenvalue mentioned above (that a large class of systems with
possible negative eigenvalues has just
 one of them was proven by Coleman\cite{co2}). \\ The factor $|\kappa|$ in
 eq.\ (27) is a dynamical one, depending on nonequilibrium properties of the
 system[3,19]. As to more recent evaluations of $ |\kappa|$ for relativistic
 systems see refs.\cite{kap}. \\ The relation between the
 entropy (19) and the exponential factor $\exp (-\beta^2/(16\pi))$ in the
 transition rate (27) associated with eq.\ (25)
  is perhaps not so obvious anymore. However,  in the case
 of nucleation one writes[3,19] \beq \Im(Z^{scl})/Z_0 =
  e^{\ds -\beta(F^{scl}-F_0)}~,~ F^{scl}=-\frac{1}{\beta}\ln \Im(Z^{scl})~,
  ~F_0
   =- \frac{1}{\beta} \ln Z_0 \eeq and interprets $\hat{F} =F^{scl}-F_0$ as
   the excess free energy of the critically large droplet.
   In our case we have $Z_0 = Z_g$ and therefore $\hat{F}_S= -(1/\beta)
   \ln Z_{GHP}$. In this sense $Z_{GHP}$ represents the partition function
   of the $bare$ black hole, i.e.\ it describes the thermodynamics of the
   black hole without the surrounding graviton gas. \\
 {\it Thus, the more sophisticated
  nucleation picture
essentially leads to the same thermodynamical properties of Schwarzschild
black holes as the "naive" use[1] of the imaginary part of the
 partition function!} \\ Still, it seems desirable to have a more systematic
 analysis of the relationship between the two approaches. \\
  Page\cite{pa1} has given a detailed analysis of the
 rate (27) for the transition of a gas into a black hole. The picture that
 black holes are formed by nucleation from a gas was already discussed by
 Gibbons and Perry\cite{gi3}.
\section{Schwarzschild black hole quantum statistics and Ising droplet
nucleation in 2 dimensions}
One of the most amazing parts of the whole analysis presented here is
that the quantum partition function (3) associated with a Schwarzschild black
hole is the same as that of the (primitive) Ising droplet model for
 nucleation
(in 1st-order phase transitions) in 2 dimensions\cite{fr}: \\ Suppose a
d-dimensional ($d \ge 2$) lattice
of $N$ Ising spins below the critical temperature $T_c$ to be in a
 positive magnetic
field $H$ so that almost all the spins are up (+1). If one then slowly turns
the magnetic field negative, the system comes into a metastable state
 in which
the total magnetization is still positive, before finally becoming negative,
too. A simple model to describe the
transition into that stable state  is the
following: Assume that (small) droplets containing $l$ down-spins form, in a
background of up-spins. The droplets are supposed to be noninteracting
(dilute-gas approximation) and the number $n_l$ of droplets of size $l$
to be given
by  \beq n_l = N e^{\ds -\beta \epsilon_l}~, l=1, \ldots~. \eeq The
droplet formation energy $\epsilon_l$ is assumed to consist of two
(competing) terms, the bulk energy $2Hl$ and a surface energy term
$\phi l^{(d-1)/d}$, so that \beq \epsilon_l = 2Hl+\phi
l^{\ds (d-1)/d}~. \eeq The interesting part of the
canonical partition function $Z_N$ is the
finite sum \beq Z_N/N = \sum_l e^{\ds -\beta 2Hl-\beta
\phi l^{ (d-1)/d}}~. \eeq
Letting $N$ go to $\infty$ and employing the grand canonical ensemble in
exactly the same way as for the instanton gas above yields the
 following grand
canonical potential $\psi$ per spin \beq \psi (\beta, H) =
 \sum_{l=0}^{\infty}e^{\ds -2l H \beta } e^{\ds -l^{(d-1)/d} \phi
 \beta}~. \eeq The theoretical investigations of the function $\psi$ have
 focussed mainly on its behaviour for negative $H$ (where the sum no longer
 converges!) in the neighbourhood of $H=0$. It is clear that $\psi$ must have
 a singularity at  $H=0$, the so-called "condensation point". Many
  approximate
analytical calculations and rigorous estimates\cite{an} suggested
 that $\psi$ has an essential
 singularity
there, a supposition supported by numerical calculations\cite{nu}. \\
Now for $d=2$ the series (32) is just the same as (3) which has been summed
exactly in ref.[1] by using Lerch's observation\cite{le} that the relation
\begin{eqnarray}
  e^{\ds -\sqrt{nx^2}}& =& \frac{|x|}{\sqrt{\pi}}\int_0^{\infty}dv
  e^{\ds -x^2v^2/4-n/v^2}\\ &=&
  \frac{|x|}{2\sqrt{\pi}}\int_0^{\infty}\frac{\ds d\tau}{\ds \tau^{3/2}}
  e^{\ds -x^2/(4\tau)-n\tau}
   \nonumber \end{eqnarray}
turns the series into a geometrical one which can be summed under the
 integral
sign and then continued analytically. We only have to put $t=-2\beta H,
x=\phi \beta$ and then get from eq.\ (6) the {\it exact} result
\beq \Im (\psi)(\beta, H<0)=
\frac{\sqrt{\pi}}{4\sqrt{2}}\phi\beta^{-1/2}|H|^{-3/2}
e^{\ds -(\phi^2 \beta)/(8|H|)}~. \eeq Thus, there is indeed an
essential singularity at $H=0$. Instead of $|H|^{-3/2}$ in front of
 the exponential
factor the approximation methods of Langer[3] and G\"unther et al.[25]
 yield the
factor $|H|$. The modified method used by Harris[26] gives the correct
$|H|$-dependence!\\ The real part of $\psi$ is to be calculated by means of
the principal value (dispersion) integral (7) (if the sum (32) starts with
$l=1$ etc.\ corresponding changes have to be made[1]). \\
In ref.[1] I pointed out that the real part (7) can become negative for
small $x$. The droplet nucleation  picture might help to understand the
(physical?) background of this. \\
For a recent discussion of metastability in the 2-dimensional
 Ising model itself see ref.\cite{ci}. \\
It was already stressed in ref.[1] that the series (3) obeys the
 heat equation
$\partial_t Z = \partial_x^2 Z.$ The corresponding partial differential
equation for $\psi$ in d=3 dimensions is $\partial_t^2 \psi = - \partial_x^3
\psi$. This ought to help analysing the behaviour of $\psi$ in the
neighbourhood of $t=0$ for d=3, too! I shall only indicate what to expect: As
that partial differential equation is invariant under the scale
 transformation $t \rightarrow \lambda t, x \rightarrow \lambda^{2/3} x$,
 there should exist solutions of the type $f(y), y=x^3/t^2$. Inserting this
 gives for $f(y)$ the ordinary diff.\ eq.\ $ 27 f'''+ (4+54/y)f''
  +6(1/y+1/y^2)f'=0$.
  For $ t \rightarrow 0, y \rightarrow \infty $, there is the approximate
 solution $f'' \sim \exp(-(4/27)y)$ which agrees with the form of
  the expected
essential singularity as to small negative $H$ in 3 dimensions[25].
 However, more analysis is certainly
 necessary here.
\section{Conclusions} It appears that there can be hardly any doubt that the
square root spectrum (1) is closely related to the expected thermodynamics
 of black
holes. It is therefore important to understand its physical meaning and the
theoretical background better. Because it implies that the area $A$
 of the horizon is proportional to $\hbar$
  one should probably look here for an interpretation[11]. This
corresponds to other recent investigations which stress the importance of
bifurcate horizons and the associated surfaces for understanding the
 thermodynamics of black
holes\cite{wa}. \\ Furthermore, the discussion above shows that a "naive" and
a droplet nucleation interpretation, respectively, of the imaginary part of a partition
function provide just  two different aspects of the thermodynamical
 properties
of  metastable systems.  \\ Similarly important is the physical understanding
of the degeneracy (2) which is equally essential for the derivation of the
thermodynamics. In the Ising droplet model above $g$ is replaced
 by the factor
$\exp{-\beta H}$ which represents the influence of the $driving$ external
(negative) magnetic field. The same role can have a
(positive) chemical potential[3,19]. Bekenstein
and Mukhanov[11] used information theoretical arguments to put $g=2$.
 However,
the arguments above go through with any $g>1$. The physical background of the
degeneracies (2) probably needs further understanding. (The "driving field"
behind $g>1$ is most likely the gravitational attraction which leads
 to the black
hole.) \\  In addition it will be interesting to see whether and how
the above considerations can be extended to the thermodynamics of the
Reissner-Nordstr{\o}m model\cite{ha4} where investigations similar to those
of refs.[8,9] for the Schwarzschild case exist\cite{th}. \\
 Very exciting is the possibility to map the quantum
statistics of a Schwarzschild black hole onto the statistics  of the
2-dimenionsal Ising model for droplet nucleation. This might allow to learn
from a known field of physics for an unknown one. \\ Finally it is surprising
that the pursuit of black hole physics leads to an exact
 mathematical solution for
an old condensed-matter problem. \\ \\ I am very much indebted
 to Malcolm Perry for
drawing my attention to refs.\ [3] and [4]. \\
{\it Note added in response to a question of the referee:} \\
If one considers a Schwarzschild black hole in d+1 space-time
dimensions\cite{ta}, $d \ge 3$, then its Schwarzschild radius $R_S$ is
proportional to $M^{1/(d-2)}$, where $M$ is the mass of the
system. Thus, if we again assume the time interval $\Delta$
in eq.\ (12) above to be proportional to $R_S$, then $\sqrt{n}$
in eq.\ (13) is replaced by $n^{(d-2)/(d-1)}$ and the resulting partition
function is the same as that of the Ising droplet model for nucleation
in d-1 dimenions (see eq.\ (32))!

 \end{document}